\documentclass[onecolumn,10pt]{article}   	
\usepackage{geometry}                		
\usepackage{graphicx}				
\usepackage{amsmath}

\providecommand{\keywords}[1]
{
  \small	
  \textbf{{Keywords---}} #1
}

\title{Gamma-ray spectra of the Crab, Vela and Geminga pulsars\\ fitted\
 with SED of the emission from their current sheet}
\author{Houshang Ardavan\footnote{ardavan@ast.cam.ac.uk}\\
Institute of Astronomy, University of Cambridge,\\
Madingley Road, Cambridge CB3 0HA, UK}
\date{}							

\begin{document}
\maketitle
\begin{abstract}
We show that the spectral energy distribution (SED) of the tightly focused radiation generated by the superluminally moving current sheet in the magnetosphere of a non-aligned neutron star fits the gamma-ray spectra of the Crab, Vela and Geminga pulsars over the entire range of photon energies so far detected by Fermi-LAT, MAGIC and HESS from them: over $10^2$ MeV to $20$ TeV.  While emblematic of any emission that entails caustics, the SED introduced here radically differs from those of the disparate emission mechanisms currently invoked in the literature to fit the data in different sections of these spectra.  We specify, moreover, the connection between the values of the fit parameters for the analysed spectra and the physical characteristics of the central neutron stars of the Crab, Vela and Geminga pulsars and their magnetospheres.
\end{abstract}


\section{Introduction}
\label{sec:introduction}

Attempts at explaining the radiation from pulsars has so far been focused mainly on mechanisms of acceleration of charged particles (see, e.g.\ the references in~\cite{Melrose2021} and~\cite{HESS2023}): an approach spurred by the fact that, once the relevant version of this mechanism is identified, one can calculate the electric current density associated with the accelerating charged particles involved and thereby evaluate the classical expression for the retarded potential that describes the looked-for radiation.  In the present paper, however, we evaluate the retarded potential, and hence the generated radiation field, using the macroscopic distribution of electric charge-current density that is already provided by the numerical computations of the structure of a non-aligned pulsar magnetosphere (section~2 of~\cite{Ardavan2021}).  Both the radiation field thus calculated and the electric and magnetic fields that pervade the pulsar magnetosphere are solutions of Maxwell's equations for the same charge-current distribution.  These two solutions are completely different, nevertheless,  because they satisfy different boundary conditions: the far-field boundary conditions with which the structure of the pulsar magnetosphere is computed are radically different from the corresponding boundary conditions with which the retarded solution of these equations (i.e. the solution describing the radiation from the charges and currents in the pulsar magnetosphere) is derived (see section~3 and the last paragraph in section~6 of~\cite{Ardavan2021}). 

Numerical computations based on the force-free and particle-in-cell formalisms have now firmly established that the magnetosphere of a non-aligned neutron star entails a current sheet outside its light cylinder whose rotating distribution pattern moves with linear speeds exceeding the speed of light in vacuum (see~\cite{SpitkovskyA:Oblique,Contopoulos:2012,Tchekhovskoy:etal} and the references in~\cite{Philippov2022}).  However, the role played by the superluminal motion of this current sheet in generating the multi-wavelength, focused pulses of radiation that we receive from neutron stars is not generally acknowledged. Given that the superluminally moving distribution pattern of this current sheet is created by the coordinated motion of aggregates of subluminally moving charged particles~\cite{GinzburgVL:vaveaa,BolotovskiiBM:VaveaD,BolotovskiiBM:Radbcm}, the motion of any of its constituent particles is too complicated to be taken into account individually.  Only the densities of charges and currents enter Maxwell's equations, on the other hand, so that the macroscopic charge-current distribution associated with the magnetospheric current sheet takes full account of the contributions toward the radiation that arise from the complicated motions of the charged particles comprising it.

The radiation field generated by a uniformly rotating volume element of the distribution pattern of the current sheet in the magnetosphere of a non-aligned neutron star embraces a synergy between the superluminal version of the field of synchrotron radiation and the vacuum version of the field of \v{C}erenkov radiation.  Once superposed to yield the emission from the entire volume of the source, the contributions from the volume elements of this distribution pattern that approach the observation point with the speed of light and zero acceleration at the retarded time interfere constructively and form caustics in certain latitudinal directions relative to the spin axis of the neutron star.  The waves that embody these caustics are more focused the farther they are from their source: as their distance from their source increases, two nearby stationary points of their phases draw closer to each other and eventually coalesce at infinity.  By virtue of their narrow peaks in the time domain, the resulting focused pulses thus procure frequency spectra whose distributions extend from radio waves to gamma-rays~\cite{Ardavan2021,ArdavanHeuristic,Ardavan2023Crab}.  

A detailed analysis of the radiation that is generated by the superluminally moving current sheet in the magnetosphere of a non-aligned neutron star can be found in~\cite{Ardavan2021}.  Heuristic accounts of the mathematical results of that analysis in more transparent physical terms are presented in~\cite{ArdavanHeuristic} and section~2 of~\cite{Ardavan2023Crab}.

We begin here by deriving the spectral energy distribution (SED) of the most tightly focused component of the radiation that is emitted by the magnetospheric current sheet (section~\ref{sec:spectrum}).  We then specify the values of the free parameters in the derived expression for which this SED best fits the data on the observed gamma-ray spectra of the Crab, Vela and Geminga pulsars (section~\ref{sec:fits}).  The specified values of the fit parameters will be used, in conjunction with the results of the analysis in~\cite{Ardavan2021}, to determine certain attributes of the central neutron stars of these pulsars and their magnetospheres in section~\ref{sec:connection}.  The radical departure of the single explanation given here for the entire breadths of the analysed spectra from those normally given in terms of disjointed sets of ad hoc spectral distribution functions (such as simple or broken power-law functions with exponential cutoffs; see, e.g.~\cite{Zanin2017}) and disparate emission mechanisms (such as synchro-curvature processes, inverse Compton scattering, or magnetic reconnection; see, e.g.~\cite{HESS2023}) is briefly discussed in the final section of the paper (section~\ref{sec:conclusion}).

\section{SED of the caustics generated by the superluminally moving current sheet}
\label{sec:spectrum}

The frequency spectrum of the radiation that is generated by the superluminally moving current sheet in the magnetosphere of a non-aligned neutron star was presented, in its general form, in section~5.3 of~\cite{Ardavan2021}.  Here we derive the SED of the most tightly-focused component of this radiation from the general expression given in eq.~(177) of that paper.  

In a case where the magnitudes of the vectors denoted by $\boldsymbol{\cal P}_l$ and $\boldsymbol{\cal Q}_l$ in eq.~(177) of~\cite{Ardavan2021} are appreciably larger than those of their counterparts, ${\bar{\boldsymbol{\cal P}}}_l$ and ${\bar{\boldsymbol{\cal Q}}}_l$, and the dominant contribution towards the Poynting flux $S_\nu$ of the radiation at the frequency $\nu$ is made by only one of the two terms corresponding to $l=1$ and $l=2$, e.g. $l=2$, eq.~(177) of~\cite{Ardavan2021} can be written as
\begin{equation}
S_\nu=\kappa_0\, k^{-2/3}\left\vert\boldsymbol{\cal P}_2\, {\rm Ai}(-k^{2/3}\sigma_{21}^2)-{\rm i}k^{-1/3}\boldsymbol{\cal Q}_2\,{\rm Ai}^\prime(-k^{2/3}\sigma_{21}^2)\right\vert^2,
\label{E1}
\end{equation}
where Ai and ${\rm Ai}^\prime$ are the Airy function and the derivative of the Airy function with respect to its argument, respectively, $k=2\pi\nu/\omega$ is the frequency $\nu$ of the radiation in units of the rotation frequency $\omega/2\pi$ of the central neutron star, and $\kappa_0$ and $\sigma_{21}$ are two positive scalars.  In the high-frequency regime $k\gg1$, the coefficients of the Airy functions in the above expression stand for the complex vectors $\boldsymbol{\cal P}_2=k^{-1/2}\boldsymbol{\cal P}_2^{(2)}$ and $\boldsymbol{\cal Q}_2=k^{-1/2}\boldsymbol{\cal Q}_2^{(2)}$ in which $\boldsymbol{\cal P}_2^{(2)}$ and $\boldsymbol{\cal Q}_2^{(2)}$ are defined by eqs.~(138)-(146) of~\cite{Ardavan2021}.  The variable $\sigma_{21}$ has a vanishingly small value at those privileged colatitudes (relative to the star's spin axis) where the high-frequency radiation is observable (section~4.5 of~\cite{Ardavan2021}).  For the purposes of the analysis in this paper, we may therefore replace $\boldsymbol{\cal P}_2^{(2)}$ and $\boldsymbol{\cal Q}_2^{(2)}$ by their limiting values for $k\gg1$ and $0\le\sigma_{21}\ll1$ and treat them as constant parameters.

Multiplying eq.~(\ref{E1}) by the radiation frequency $\nu=\omega k/2\pi$ and expressing $\boldsymbol{\cal P}_2$ and $\boldsymbol{\cal Q}_2$ in terms of $\boldsymbol{\cal P}_2^{(2)}$ and $\boldsymbol{\cal Q}_2^{(2)}$, we find that the SED of this emission is given by
\begin{equation}
\nu S_\nu=\frac{\omega\kappa_0}{2\pi}k^{-2/3}\left\vert\boldsymbol{\cal P}^{(2)}_2\, {\rm Ai}(-k^{2/3}\sigma_{21}^2)
-{\rm i}k^{-1/3}\boldsymbol{\cal Q}^{(2)}_2\,{\rm Ai}^\prime(-k^{2/3}\sigma_{21}^2)\right\vert^2.
\label{E2}
\end{equation}
Evaluation of the right-hand side of eq.~(\ref{E2}) results in
\begin{eqnarray}
\nu S_\nu&=&\kappa_1\, k^{-2/3}\left[{\rm Ai}^2(-k^{2/3}\sigma_{21}^2)+\zeta_1^2k^{-2/3}{{\rm Ai}^\prime}^2(-k^{2/3}\sigma_{21}^2)\right.\nonumber\\*
&&\left.+2\zeta_1\cos\beta\, k^{-1/3}{\rm Ai}(-k^{2/3}\sigma_{21}^2){\rm Ai}^\prime(-k^{2/3}\sigma_{21}^2)\right],
\label{E3}
\end{eqnarray}
where 
\begin{equation}
\kappa_1=\frac{\omega\kappa_0}{2\pi} \left\vert\boldsymbol{\cal P}^{(2)}_2\right\vert^2,\quad \zeta_1=\frac{\left\vert{\boldsymbol{\cal Q}^{(2)}_2}\right\vert}{\left\vert{\boldsymbol{\cal P}^{(2)}_2}\right\vert},\quad\cos\beta=\frac{\Im\left(\boldsymbol{\cal Q}^{(2)}_2\cdot\boldsymbol{\cal P}^{(2)*}_2\right)}{\left\vert\boldsymbol{\cal Q}^{(2)}_2\right\vert \left\vert\boldsymbol{\cal P}^{(2)}_2\right\vert},
\label{E4}
\end{equation}
and $\Im{}$ and $*$ denote an imaginary part and the complex conjugate, respectively.  The above spectrum is emblematic of any radiation that entails caustics (see~\cite{Stamnes1986}).

To take account of the fact that the parameter $\sigma_{21}$ assumes a non-zero range of values across the (non-zero) latitudinal width of the detected radiation beam (section~4.5 of~\cite{Ardavan2021}), we must integrate $\nu S_\nu$ with respect to $\sigma_{21}$ over a finite interval $\rho\sigma_0\le\sigma_{21}\le\sigma_0$ with $\sigma_0\ll1$ and $0\le\rho<1$.  Performing the integration of the Airy functions in eq.~(\ref{E3}) with respect to $\sigma_{21}$ by means of Mathematica, we thus obtain 
\begin{equation}
{\cal S}_\nu=\nu\int_{\rho\sigma_0}^{\sigma_0}S_\nu\,{\rm d}\sigma_{21}
= \kappa\left[f_1(\chi,\rho)+\frac{\zeta^2}{4\sqrt{3}}f_2(\chi,\rho)-\frac{\zeta\cos\beta}{2\sqrt{3}}f_3(\chi,\rho)\right],
\label{E5}
\end{equation}
where
\begin{equation}
\kappa=\left(\frac{\sigma_0}{\sqrt{3\pi}}\right)^3\kappa_1,\qquad \zeta=\sigma_0 \zeta_1,\qquad\chi=\frac{2}{3}{\sigma_0}^3k,
\label{E6}
\end{equation}
\begin{eqnarray}
f_1&=&\bigg[3\Gamma\left(\frac{7}{6}\right)\eta\chi^{-2/3}{}_2F_3\left(\begin{matrix}1/6&1/6&{}\\1/3&2/3&7/6
\end{matrix}
;-\eta^6\chi^2
\right)+\pi^{1/2}\eta^3{}_2F_3
\left(\begin{matrix}
1/2&1/2&{}\\
2/3&4/3&3/2
\end{matrix}
\,;-\eta^6\chi^2
\right)\nonumber\\
&&+\frac{9}{20}\Gamma\left(\frac{5}{6}\right)\eta^5\chi^{2/3}{}_2F_3
\left(\begin{matrix}
5/6&5/6&{}\\
4/3&5/3&11/6
\end{matrix}
\,;-\eta^6\chi^2
\right)\bigg]_{\eta=\rho}^{\eta=1},\nonumber\\
\label{E7}
\end{eqnarray}
\begin{equation}
f_2=\eta\chi^{-4/3}\,{}_{24}G^{31}\left(-\eta^2\chi^{2/3},\frac{1}{3}\left\vert\,\begin{matrix}
5/6&7/6&{}&{}\\
0&2/3&4/3&-1/6
\end{matrix}\right)\right.\bigg\vert_{\eta=\rho}^{\eta=1},
\label{E8}
\end{equation}
\begin{equation}
f_3=\eta\chi^{-1}\,{}_{24}G^{31}\left(-\eta^2\chi^{2/3},\frac{1}{3}\left\vert\,\begin{matrix}
5/6&1/2&{}&{}\\
0&1/3&2/3&-1/6
\end{matrix}\right)\right.\bigg\vert_{\eta=\rho}^{\eta=1},
\label{E9}
\end{equation}
and ${}_2F_3$ and ${}_{24}G^{31}$ are respectively the generalised hypergeometric function~\cite{Olver} and the generalised Meijer G-Function.\footnote{ {{https://mathworld.wolfram.com/Meijer G-Function.html}}}  The variable $\chi$ that appears in the above expressions is related to the frequency $\nu$ of the radiation via $\chi=4\pi{\sigma_0}^3\nu/(3\omega)$.

The scale and shape of the SED given in eq.~(\ref{E5}) depend on the five parameters $\kappa$, $\zeta$, $\beta$, $\sigma_0$ and $\rho$: parameters whose values are dictated by the characteristics of the magnetospheric current sheet (see section~\ref{sec:connection}).  The parameters $\zeta$, $\beta$  and $\rho$ determine the shape of the energy distribution while the parameters $\kappa$ and $\sigma_0$ determine the position of this distribution along the energy-flux and photon-energy axes, respectively.

\begin{figure}
\centerline{\includegraphics[width=12cm]{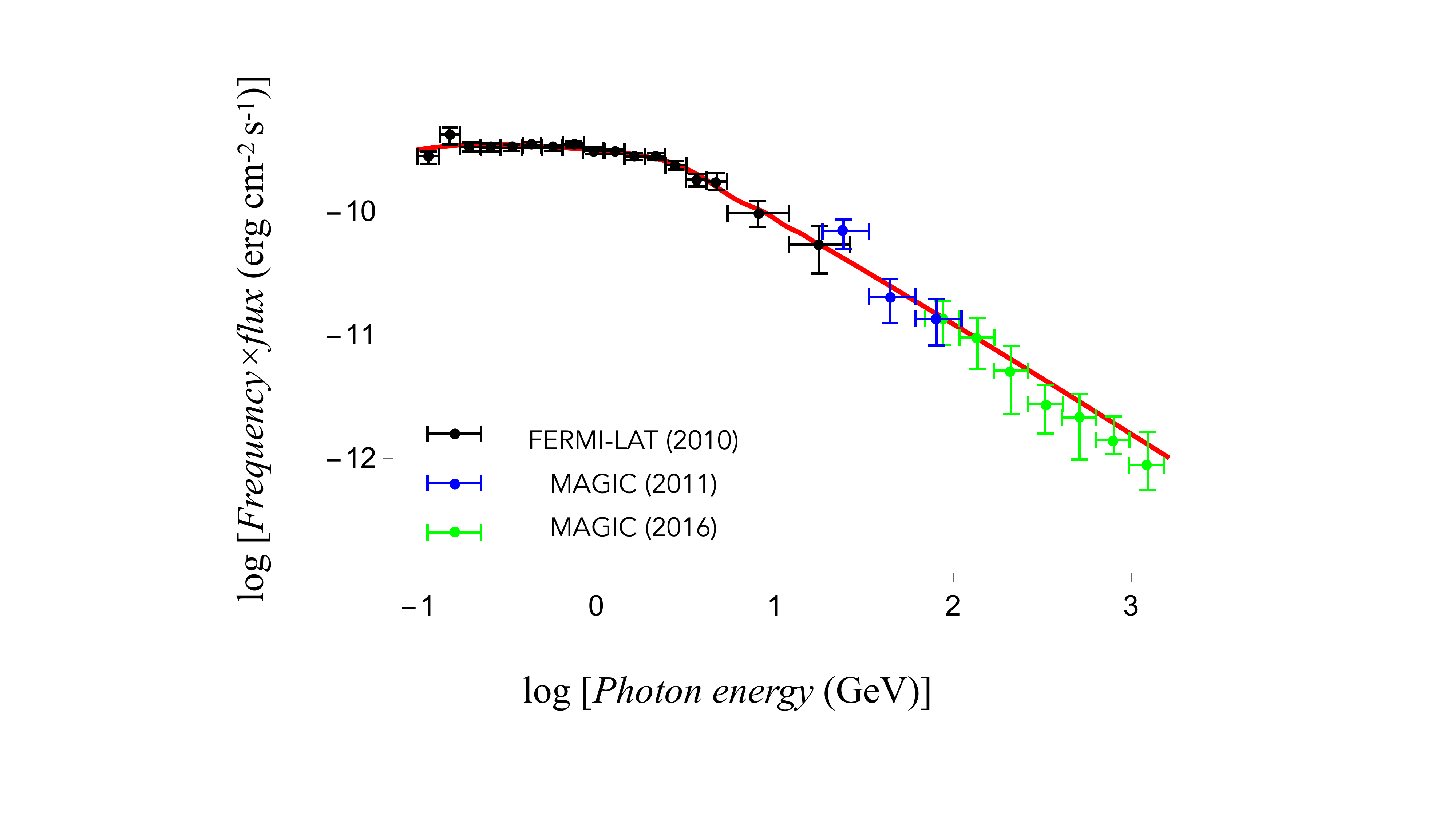}}
\caption{Phase-averaged gamma-ray spectrum of the Crab pulsar.  The Fermi-LAT data points below $25$ GeV (coloured black) are those reported in~\cite{Abdo2010}.  The MAGIC data points between $25$ and $100$ GeV (coloured blue) are those reported in~\cite{Aleksic2011}.  The MAGIC data points beyond $100$ Gev (coloured green) are extracted, by the procedure described in section~4 of~\cite{Ardavan2023Crab}, from those reported by~\cite{Ansoldi2016}.  The curve (coloured red) is a plot of the SED described by eq.~(\ref{E5}) for the parameters given in eq.~(\ref{E10}).}
\label{CVG1}
\end{figure}

\section{Fits to the gamma-ray data on the SEDs of the Crab, Vela and Geminga pulsars}
\label{sec:fits}

\begin{figure}
\centerline{\includegraphics[width=12cm]{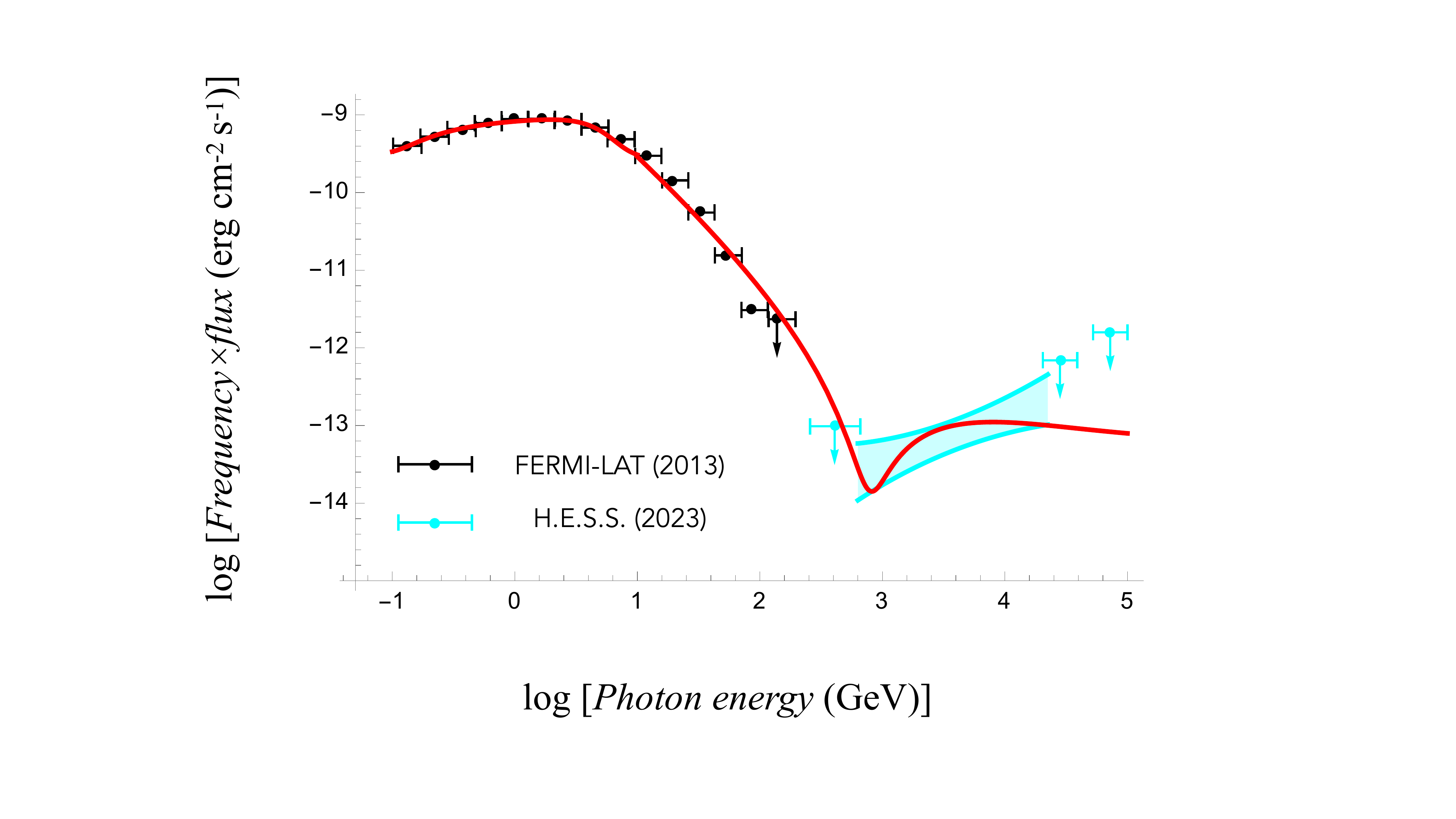}}
\caption{Gamma-ray spectrum of the Vela pulsar.  The Fermi-LAT data points below $10^2$ GeV (coloured black) are those reported in~\cite{Abdo2013}.  The HESS data above $260$ GeV (shown as the magenta area between $1\sigma$ uncertainty bands and the magenta upper limits) are those reported in~\cite{HESS2023}.  The curve (coloured red) is a plot of the SED described by eq.~(\ref{E5}) for the parameters given in eqs.~(\ref{E11}) and (\ref{E12}).}
\label{CVG2}
\end{figure}

In this section we use Mathematica's {\tt `NonlinearModelFit'} procedure\footnote{ {{https://reference.wolfram.com/language/ref/NonlinearModelFit.html}}} and the statistical information that it provides to determine the values of the fit parameters and their standard errors.  Where, owing to the complexity of the expression in eq.~(\ref{E5}), this procedure fails to work and so the fits to the data are obtained by elementary iteration, only the values of these parameters are specified.  The fit residuals in the case of each pulsar turn out to be smaller than the corresponding observational errors for almost all data points (see figures~\ref{CVG1}--\ref{CVG3}).  Since there are no two values of any of the fit parameters for which the present SED has the same shape and position, the specified values of the fit parameters are moreover unique.

\subsection{Spectrum of the Crab pulsar (PSR J0534+2200)}
\label{subsec:Crab}

Figure~\ref{CVG1} shows the data on the energy spectrum of the phase-averaged gamma-ray emission from the Crab pulsar~\cite{Abdo2010, Aleksic2011, Ansoldi2016} and a plot of the SED described by eq.~(\ref{E5}) that fits them best.  The parameters for which the function ${\cal S}_\nu(\chi)$ (depicted by the red curve in this figure) is plotted have the following values and standard errors:
\begin{eqnarray}
\kappa&=&(8.15\pm0.75)\times10^{-11}\quad{\rm erg}\,{\rm cm}^{-2}\,{\rm s}^{-1},\nonumber\\*
\chi&=&(0.576\pm0.095)\times(h\nu)_{\rm GeV},\nonumber\\*
\sigma_0&=&(4.76\pm0.28)\times10^{-8},\,\,\, \zeta=0.394\pm0.076,\nonumber\\*
\beta&=&0.258\pm0.249,\quad\rho=0,
\label{E10}
\end{eqnarray}
in which $h$ and $(h\nu)_{\rm GeV}$ stand for the Planck constant and photon energy in units of GeV, respectively, and $2\pi/\omega$ has been set equal to the period of the Crab pulsar ($0.033$ s).  (Figure~\ref{CVG1} was first reported in~\cite{Ardavan2023Crab} but with an incorrect label on its vertical axis.)

As the photon energy $h\nu$ increases beyond $1.2$ TeV, the slope of the curve in figure~\ref{CVG1} continues to decrease at the relatively slow rate in which it decreases past $50$ GeV.  Thus the upper limits given in~\cite{Ansoldi2016} for the flux density of gamma-rays whose energies exceed $1.2$ TeV all lie above the continuation of the curve shown in figure~\ref{CVG1}.

\subsection{Spectrum of the Vela pulsar (PSR J0835-4510)}
\label{subsec:Vela} 

The data on the Vela pulsar both in the GeV band detected by Fermi-LAT~\cite{Abdo2013} and in the TeV band detected by HESS~\cite{HESS2023} are shown in figure~\ref{CVG2}.  The SED that best fits these data (represented by the red curve in figure~\ref{CVG2}) is described by the expression in eq.~(\ref{E5}) for the following values of its free parameters:
\begin{eqnarray}
\kappa&=&(3.31\pm0.047)\times10^{-10}\quad{\rm erg}\,{\rm cm}^{-2}\,{\rm s}^{-1},\nonumber\\*
\chi&=&(0.436\pm0.071)\times(h\nu)_{\rm GeV},\nonumber\\*
\sigma_0&=&(3.12\pm0.18)\times10^{-8},\quad \zeta=0.518\pm0.065,\nonumber\\*
\beta&=&0.244\pm0.073,\quad\rho=0.409\pm0.198,
\label{E11}
\end{eqnarray}
over the range $0.1\le(h\nu)_{\rm GeV}\le10$ and 
\begin{eqnarray}
\kappa&=&7.94\times10^{-13}\quad{\rm erg}\,{\rm cm}^{-2}\,{\rm s}^{-1},\nonumber\\*
\chi&=&2.51\times10^{-6}\times(h\nu)_{\rm GeV},\nonumber\\*
\sigma_0&=&5.59\times10^{-10},\,\, \zeta=0.203,\,\,\beta=0.105,\,\,\rho=0.99,
\label{E12}
\end{eqnarray}
over the range $10\le(h\nu)_{\rm GeV}\le2\times10^4$ of photon energies.

The difference between the values of the free parameters in the above two ranges of photon energies stems from the fact that the higher-frequency radiation is appreciably more focused than its lower-frequency counterpart in this case.  This is reflected not only in the much lower value of $\sigma_0$ in eq.~(\ref{E12}) compared to that in eq.~(\ref{E11}), but also in the higher value of $\rho$ (i.e.\ the shorter range of values of $\sigma_{21}$): the smaller the value of $\sigma_{21}$, the closer to each other are the stationary points of the phases of the emitted waves and so the more focused is the radiation (see section~4.5 of~\cite{Ardavan2021}).

\begin{figure}
\centerline{\includegraphics[width=12cm]{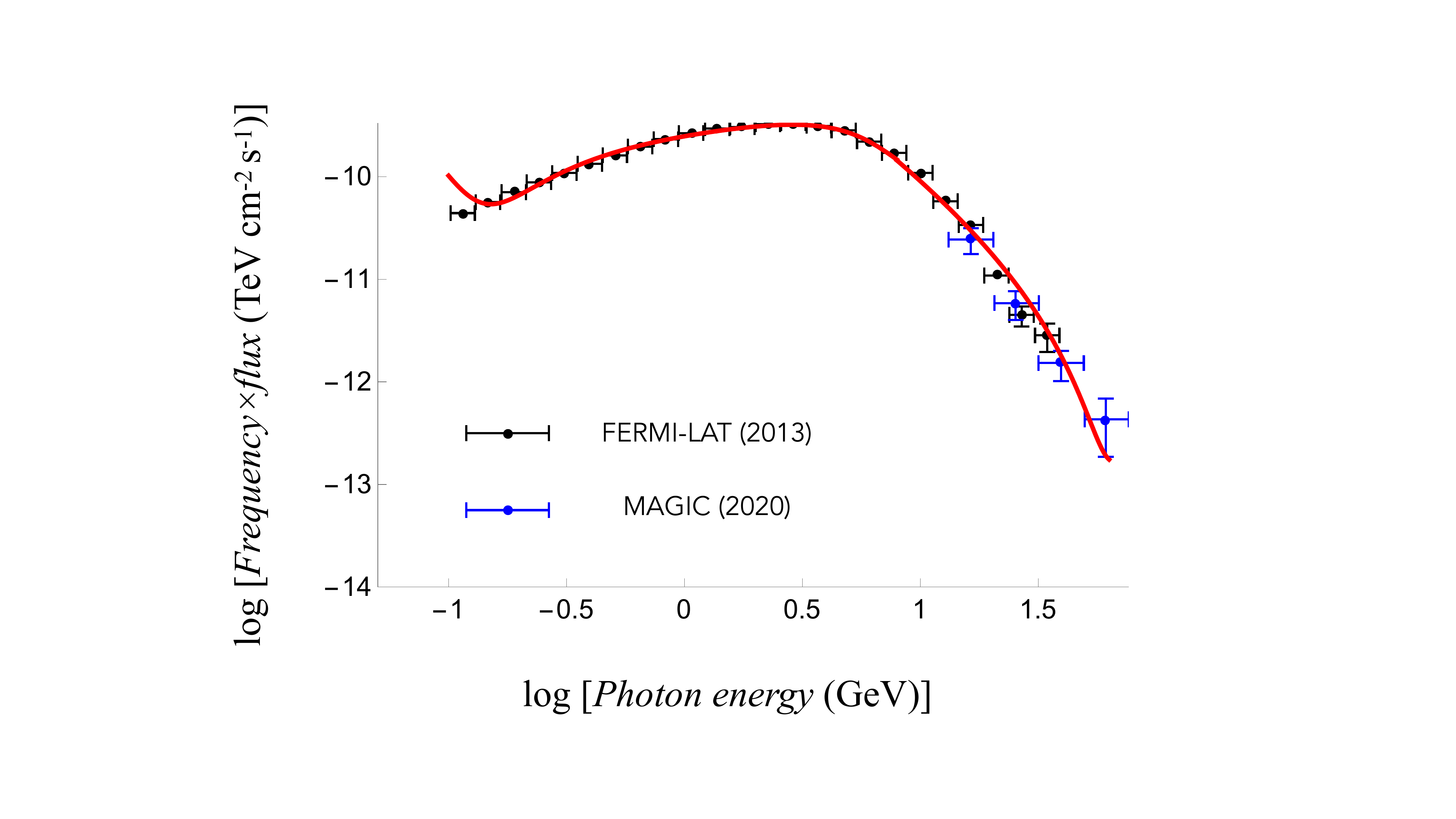}}
\caption{Gamma-ray spectrum of the Geminga pulsar.  The Fermi-LAT data points below $35$~GeV (coloured black) are those reported in~\cite{Abdo2013}.  The MAGIC data points above $35$~GeV (coloured blue) are those reported in~\cite{MAGIC2020}.  The curve (coloured red) is a plot of the SED described by eq.~(\ref{E5}) for the parameters given in eqs.~(\ref{E13}) and (\ref{E14}).}
\label{CVG3}
\end{figure}

\subsection{Spectrum of the Geminga pulsar (PSR J0633+1746)} 
\label{sebsec:Geminga}

The Fermi-LAT~\cite{Abdo2013} and MAGIC~\cite{MAGIC2020} data on the gamma-ray emission from the Geminga pulsar are shown in figure~\ref{CVG3}.  The fit to these data (the red curve in figure~\ref{CVG3}) is described by the expression in eq.~(\ref{E5}) for 
\begin{eqnarray}
\kappa&=&(1.47\pm0.23)\times10^{-10}\quad{\rm TeV}\,{\rm cm}^{-2}\,{\rm s}^{-1},\nonumber\\*
\chi&=&(0.392\pm0.038)\times(h\nu)_{\rm GeV},\nonumber\\*
\sigma_0&=&(2.17\pm0.07)\times10^{-8},\quad \zeta=0.646\pm0.028\nonumber\\*
\beta&=&0.194\pm0.017,\quad\rho=0.568\pm0.131,
\label{E13}
\end{eqnarray}
over the range $0.1\le(h\nu)_{\rm GeV}\le7.94$ and 
\begin{eqnarray}
\kappa&=&1.01\times10^{-13}\quad{\rm TeV}\,{\rm cm}^{-2}\,{\rm s}^{-1},\nonumber\\*
\chi&=&3.98\times10^{-6}\times(h\nu)_{\rm GeV},\nonumber\\*
\sigma_0&=&4.71\times10^{-10},\,\, \zeta=0.101,\,\,\beta=0.071,\,\,\rho=0.511,
\label{E14}
\end{eqnarray}
over the range $7.94\le(h\nu)_{\rm GeV}\le63.1$ of photon energies.  As in the case of the spectrum of the Vela pulsar, the change in the values of the fit parameters across $(h\nu)_{\rm GeV}=7.94$ reflects the fact that the higher-frequency radiation is appreciably more focused than its lower-frequency counterpart.

\section{The connection between the parameters of the fitted spectra and the physical characteristics of their sources}
\label{sec:connection}

From eqs.~(\ref{E6}), (\ref{E4}) and (\ref{E1}), it follows that the parameters $\kappa$, $\zeta$, $\rho$ and $\sigma_0$ in eq.~(\ref{E5}) are related to the characteristics of the source of the observed radiation via the quantities $\omega$, $\kappa_0$, $\boldsymbol{\cal P}_2^{(2)}$ and $\boldsymbol{\cal Q}_2^{(2)}$ that appear in the expression for the SED ${\cal S}_\nu$.  In this section we use the results of the analysis presented in~\cite{Ardavan2021} to express these quantities in terms of the inclination angle of the central neutron star, $\alpha$, the magnitude of the star's magnetic field at its magnetic pole, $B_0=10^{12}{\hat B}_0$ Gauss, the radius of the star, $r_{s0}=10^6 d$ cm, the rotation frequency of the star, $\omega=10^2{\hat P}^{-1}$ rad/s, and the spherical polar coordinates, $R_P=D\,\,{\rm kpc}=3.085\times10^{21}D$~cm, $\varphi_P$ and $\theta_P$, of the observation point $P$ in a frame whose centre and $z$-axis coincide with the centre and spin axis of the star.

When $\sigma_{21}\ll1$ and so the quantities $\tau_{2{\rm min}}$ and $\tau_{2{\rm max}}$ that appear in eqs.~(177), (138)--(146) and (136) of~\cite{Ardavan2021} are approximately equal, these equations imply that
\begin{equation}
\kappa_0=4.15\times10^{18} w_1^2 w_3^2 {\hat B}_0^2 d^4 D^{-2} {\hat P}^{-1}\quad {\rm Jy},
\label{E15}
\end{equation}
and
\begin{equation}
\left\vert\boldsymbol{\cal P}_2^{(2)}\right\vert=\frac{b}{3}\left(\frac{2}{\pi a^3}\right)^{1/2}\left(\frac{2\sigma_{21}}{\partial^2 f_{2C}/\partial\tau^2|_{\tau=\tau_{2{\rm min}}}}\right)^{1/2}\left\vert{\tilde{\bf P}}_2\right\vert,
\label{E16}
\end{equation}
where
\begin{equation}
w_1=\left\vert1-2\alpha/\pi\right\vert,\qquad w_3=1+0.2\sin^2\alpha,
\label{E17}
\end{equation}
\begin{eqnarray}
f_{2C}&=&({\hat r}_P^2{\hat r}_{sC}^2\sin^2\theta-1)^{1/2}-{\hat R}_P-\arccos\left({\hat r}_P^{-1}{\hat r}_{sC}^{-1}\csc\theta\right)\nonumber\\*
&&-\arccos(\cot\alpha\cot\theta)+{\hat r}_{sC}+\varphi_P-{\hat r}_{s0}-2\pi,
\label{E18}
\end{eqnarray}
\begin{equation}
a=({\hat r}_P^2{\hat r}_{sC}^2\sin^2\theta-1)\left[\frac{({\hat r}_P^2{\hat r}_{sC}^2\sin^2\theta-1)^{1/2}+{\hat r}_{sC}}{{\hat r}_{sC}({\hat r}_P^2-1)^{1/2}({\hat R}_P^2\sin^2\theta-1)^{1/2}}-\frac{1}{({\hat r}_P^2{\hat r}_{sC}^2\sin^2\theta-1)^{1/2}}\right],
\label{E19}
\end{equation}
\begin{equation}
b=\frac{{\hat r}_P^2{\hat r}_{sC}^2\sin^2\theta-1}{{\hat r}_{sC}({\hat r}_P^2-1)^{1/2}({\hat R}_P^2\sin^2\theta-1)^{1/2}},
\label{E20}
\end{equation}
\begin{equation}
{\hat r}_{sC}=\frac{({\hat r}_P^2-1)^{1/2}({\hat R}_P^2\sin^2\theta-1)^{1/2}-{\hat z}_P\cos\theta}{{\hat r}_P^2\sin^2\theta-1},
\label{E21}
\end{equation}
\begin{equation}
\sigma_{21}=[{\textstyle\frac{3}{4}}(f_{2C}\vert_{\tau=\tau_{2{\rm max}}}-f_{2C}\vert_{\tau=\tau_{2{\rm min}}})]^{1/3},
\label{E22}
\end{equation}
\begin{equation}
\theta=\arccos\left(\sin\alpha\cos\tau\right),
\label{E23}
\end{equation}
\begin{eqnarray}
\left\vert{\tilde{\bf P}}_2\right\vert &=&\cos\alpha[{\hat r}_{sC}^2(1+\cos^2\theta_P-2\cos\theta\cos\theta_P)+\tan^2\alpha\nonumber\\*
&&-\cot^2\theta+2{\hat r}_{sC}\sin\theta\cos\theta_P(\tan^2\alpha-\cot^2\theta)^{1/2}]^{1/2},
\label{E24}
\end{eqnarray}
and the variables $\tau_{2{\rm min}}$ and $\tau_{2{\rm max}}$ stand for the minimum and maximum of the function $f_{2C}$ (see also eqs.~(7), (9), (174), (175), (93)--(95), (97) and (88) of~\cite{Ardavan2021}).  The caret on $R_P$ and $r_{s0}$ (and $r_P=R_P\sin\theta_P$, $z_P=R_P\cos\theta_P$) is used here to designate a variable that is rendered dimensionless by being measured in units of the light-cylinder radius $c/\omega$.  (Note the following two corrections: the vector ${\bf P}_2$ in eq.~(145) and the numerical coefficient $1.54\times10^{19}$ in eq.~(177) of~\cite{Ardavan2021} have been corrected to read ${\tilde{\bf P}}_2$ and $4.15\times10^{18}$, respectively.)

The expression for $\vert{\tilde{\bf P}}_2\vert$ in eq.~(\ref{E24}) is derived from that for ${\bf P}_2$ given by eqs.~(98), (80), (78) and (62) of~\cite{Ardavan2021}.  In this derivation, we have set the observation point on the cusp locus of the bifurcation surface where ${\hat r}_s={\hat r}_{sC}$, have approximated $(p_1,p_2,p_3)$ by its far-field value $2^{1/3}{\hat R}_P^{-1}({\hat R}_P^{-1}, -1, 1)$ and have let ${\bf P}_2={\hat R}_P^{-1}{\tilde{\bf P}}_2$.  The factor ${\hat R}_P^{-2}$ that would have otherwise appeared in the resulting expression for $\vert\boldsymbol{\cal P}^{(2)}_2\vert^2$ is thus incorporated in the factor $\kappa_0$ in eq.~(\ref{E15}).

\begin{figure}
\centerline{\includegraphics[width=10cm]{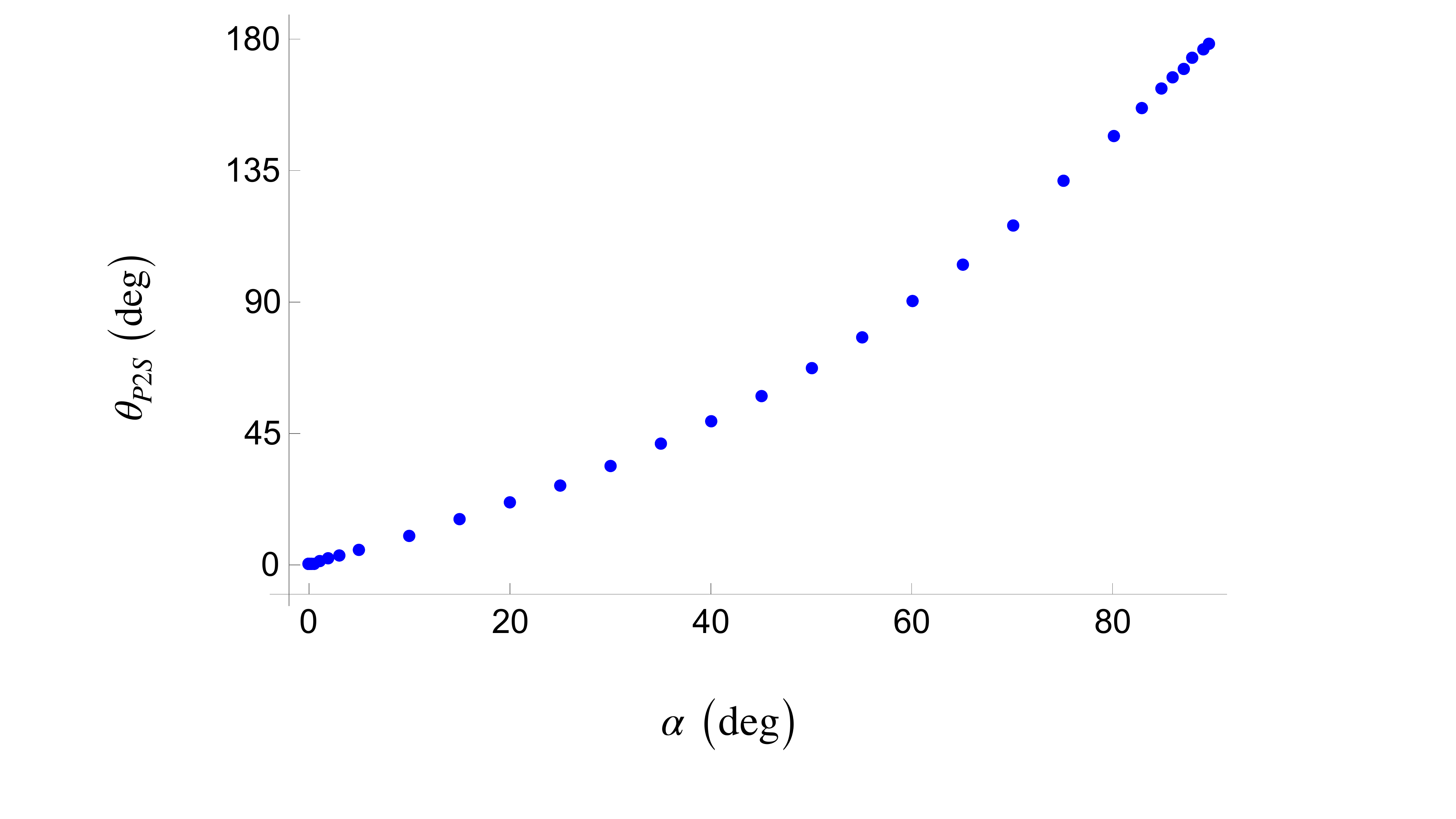}}
\caption{The dependence on the star's inclination angle $\alpha$ of the critical colatitude $\theta_{P2S}$ along which the high-frequency radiation beam generated by the superluminally moving current sheet propagates.}
\label{CVG4}
\end{figure}

For certain values of $\theta_P$, denoted by $\theta_{P2S}$, the function $f_{2C}(\tau)$ has an inflection point~(see section 4.4 of~\cite{Ardavan2021}, and section 2 of~\cite{Ardavan2023Crab}).  For any given inclination angle $\alpha$, the position $\tau_{2S}$ of this inflection point and the colatitude $\theta_{P2S}$ of the observation points for which $f_{2C}(\tau)$ has an inflection point follow from the solutions to the simultaneous equations $\partial f_{2C}/\partial\tau=0$ and $\partial^2 f_{2C}/\partial\tau^2=0$.  (Explicit expressions for the derivatives that appear in these equations can be found in appendix A of~\cite{Ardavan2021}.)   For values of $\theta_P$ close to $\theta_{P2S}$, the separation between the maximum $\tau=\tau_{2{\rm max}}$ and minimum $\tau=\tau_{2{\rm min}}$ of $f_{2C}$ and hence the value of $\sigma_{21}$ are vanishingly small.  Since the high-frequency radiation that arises from the current sheet is detectable only in the vicinity of the latitudinal direction $\theta_P=\theta_{P2S}$ (see figure~\ref{CVG4}), here we have evaluated the variables that appear in eqs.~(\ref{E16})--(\ref{E24}) for $\sigma_{21}\ll1$.  The ratio appearing inside the second pair of parentheses in eq.~(\ref{E16}), which becomes indeterminate as $\sigma_{21}$ tends to zero, approaches a finite value in this limit.  

The dimensionless frequency $k=2\pi\nu/\omega$ of a $1$ GeV photon has the value $8\times10^{21}$ in the case of the Crab pulsar.  In contrast, as indicated by the small values of the fit parameter $\sigma_0$ encountered in section~\ref{sec:fits}, the value of the variable $\chi=2\sigma_0^3k/3$ that appears in the description of the SED ${\cal S}_\nu$ is of the order of $10^{-3}$ for a $1$ GeV gamma-ray photon.  Because the measurement of the spectrum described by eq.~(\ref{E5}) is carried out in the cases considered here by counting the number of detected photons whose energies lie between $10^{-1}$ to $\sim10^3$ GeV, the widths of the equivalent frequency bins over which each value of the flux is determined are given by $\Delta k=3\sigma_0^{-3}\Delta\chi/2$ with a $\Delta\chi$ whose values correspondingly range from $\sim10^{-4}$ to $\sim1$. 

Once the the expression for the SED ${\cal S}_\nu$ is accordingly multiplied by $3\sigma_0^{-3}\Delta\chi/2$, eqs.~(\ref{E6}), (\ref{E4}), (\ref{E15}) and (\ref{E16}) jointly yield
\begin{equation}
\kappa=2.42\times10^{-6} {\hat B}_0^2d^4D^{-2}{\hat P}^{-2}\Delta\chi\,{\hat\kappa}_{\rm th}\qquad {\rm erg}\,\,{\rm s}^{-1}{\rm cm}^{-2},
\label{E25}
\end{equation}
in which 
\begin{equation}
{\hat\kappa}_{\rm th}=w_1^2 w_3^2\frac{b^2}{a^3}\left(\frac{2 \sigma_{21}} {\partial^2 f_{2C}/\partial\tau^2|_{\tau=\tau_{2{\rm min}}}}\right)\left\vert{\tilde{\bf P}}_2\right\vert^2,
\label{E26}
\end{equation}
and Jy has been expressed in terms of erg s${}^{-1}$ cm${}^{-2}$ Hz${}^{-1}$.  Equation~(\ref{E25}) can be written as ${\hat\kappa}_{\rm th}={\hat\kappa}_{\rm obs}$, where
\begin{equation}
{\hat\kappa}_{\rm obs}=4.13\times10^5({\hat B}_0d^2)^{-2}(\Delta\chi)^{-1}D^2{\hat P}^2\kappa.
\label{E27}
\end{equation}
While ${\hat\kappa}_{\rm obs}$ only contains the observed parameters of the pulsar and its emission, the value of ${\hat\kappa}_{\rm th}$ is determined by the physical characteristics of the magnetospheric current sheet that acts as the source of the observed emission.  When, as in figure~20 of~\cite{Ardavan2021}, the value of the colatitude $\theta_P$ of the observation point is sufficiently close to that of the critical angle $\theta_{P2S}$ for $\vert\tau_{2{\rm max}}-\tau_{2{\rm min}}\vert$ and hence $\sigma_{21}$ to be much smaller than unity, the right-hand side of eq.~(\ref{E26}) is a function of the inclination angle $\alpha$ and the observer's distance ${\hat R}_P$ only. 

Equation~(\ref{E25}) thus enables us to connect the parameters of the fitted spectra to the physical characteristics of their sources.

\begin{figure}
\centerline{\includegraphics[width=10cm]{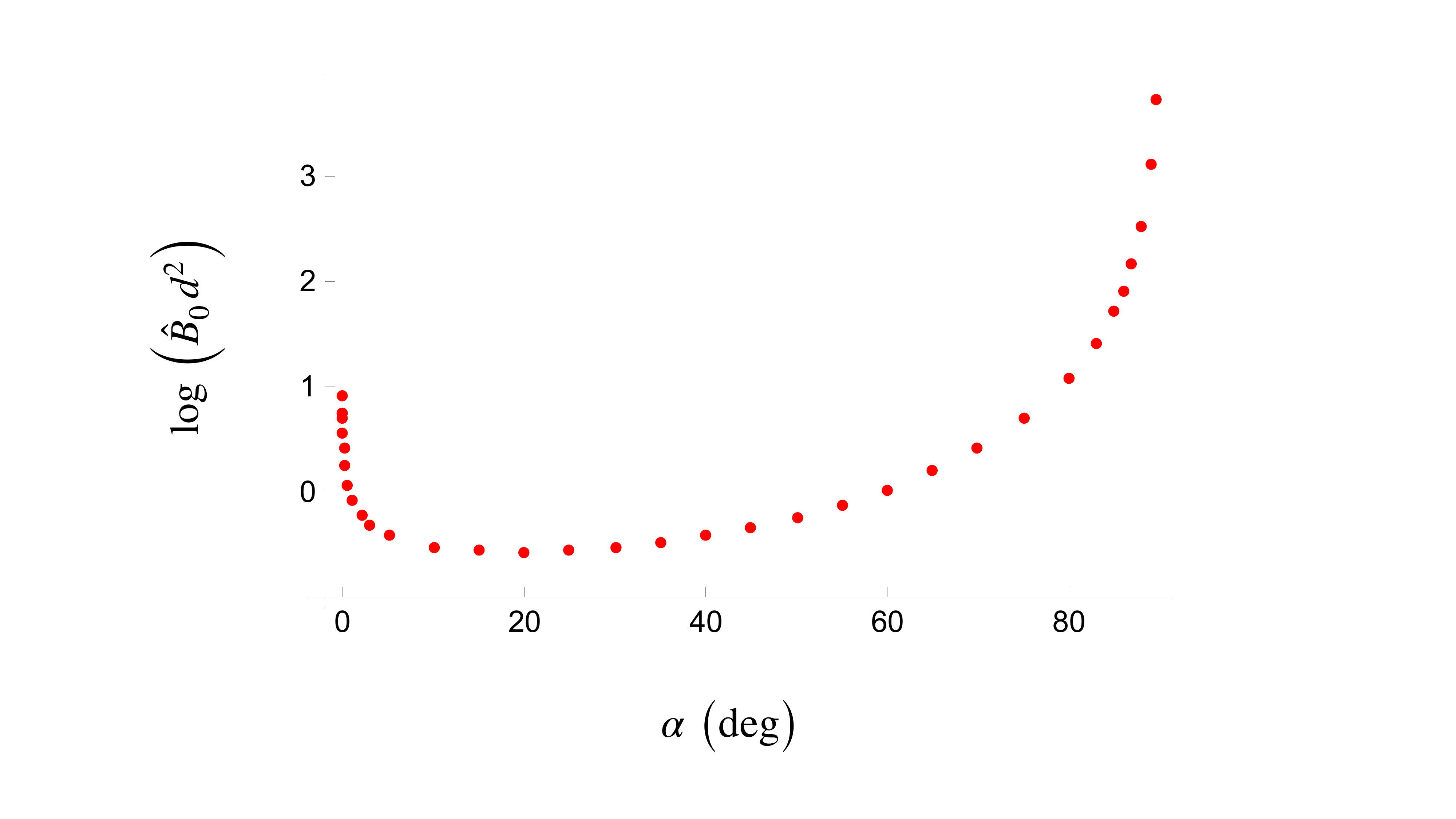}}
\caption{Dependence of the ${\hat B}_0 d^2$ on $\alpha$ for the Crab pulsar, where ${\hat B}_0$ is the magnitude of the star's magnetic field at its magnetic pole in units of $10^{12}$ Gauss, $d$ is the radius of the star in units of $10^6$ cm and $\alpha$ is the angle between the magnetic and rotation axes of the star in degrees.}
\label{CVG5}
\end{figure}

\subsection{Characteristics of the source of the Crab pulsar's emission}
\label{subsec:CrabSource}

Once the value of the fit parameter $\kappa$ given in eq.~(\ref{E10}), the period ($0.033$ s) and the distance ($2$ kpc) of the Crab pulsar are inserted in eq.~(\ref{E27}), the resulting value of  ${\hat\kappa}_{\rm obs}$ can be equated to ${\hat\kappa}_{\rm th}$ to obtain
\begin{equation}
{\hat B}_0d^2=1.93\times10^{-1} {\hat\kappa}_{\rm th}^{-1/2}
\label{E28}
\end{equation}
for $\Delta\chi=10^{-3}$.  The dependence of ${\hat B}_0d^2$ on $\alpha$ implied by this relationship, in conjunction with eq.~(\ref{E26}), is plotted in figure~\ref{CVG5}.

One can infer from the separation between the two peaks in the gamma-ray light curve of the Crab pulsar ($\sim 0.4$ cycles~\cite{Abdo2010}) that the inclination angle $\alpha$ of this pulsar has a value lying in the range $60^\circ$ to $70^\circ$ (see section 5.1 of~\cite{Ardavan2021}).  This in turn implies that the value of ${\hat B}_0d^2$ lies between $1.05$ and $2.61$ in this case (see figure~\ref{CVG5}), i.e.\ that the magnitude of the star's magnetic field at its magnetic pole lies in the range $1.05\times10^{12}$ to $2.61\times10^{12}$ Gauss if we assume that the star's radius has the value $10^6$ cm.  These are of the same order of magnitude as the value of $B_0$ obtained from the conventional formula for magnetic dipole radiation (i.e.\ $3.79\times10^{12}$ Gauss~\cite{Manchester2005}).

The corresponding value of the privileged latitudinal direction $\theta_{P2S}$ in which the high-frequency radiation from the Crab pulsar can be observed lies between $90^\circ$ and $117^\circ$ according to figure~\ref{CVG4}.  The range of values of the inclination angle $\alpha$ determines also the scale factor $j_0$ of the electric current density in the magnetosphere of the pulsar: from eq.~(22) of~\cite{Ardavan2021} it follows that
\begin{equation}
j_0=1.29\times10^8{\hat B}_0d^2{\hat P}^{-3}w_1w_3,\qquad {\rm statamp\,cm}^{-2}
\label{E29}
\end{equation}
a scale factor whose value lies between $3.55\times10^8$ and $6.01\times10^8$ statamp cm$^{-2}$ in the case of the Crab pulsar.  

The above results supersede those reported in section 4 of~\cite{Ardavan2023Crab}.

\subsection{Characteristics of the source of the Vela pulsar's emission}
\label{subsec:VelaSource}

Properties of the central neutron star of the Vela pulsar and its magnetosphere are more accurately reflected in the parameters of its spectrum over the photon energies $0.1$ to $10$ GeV, i.e.\ the interval that encompasses the majority of the data points plotted in figure~\ref{CVG2}.  Insertion of the value of the fit parameter $\kappa$ given in eq.~(\ref{E11}), the period ($0.089$~s) and the distance ($287$~pc) of this pulsar in eq.~(\ref{E27}), together with ${\hat\kappa}_{\rm obs}={\hat\kappa}_{\rm th}$, results in
\begin{equation}
{\hat B}_0d^2=1.50\times10^{-1} {\hat\kappa}_{\rm th}^{-1/2}
\label{E30}
\end{equation}
for $\Delta\chi=10^{-3}$.  Thus the dependence of ${\hat B}_0d^2$ on $\alpha$ is given by a shifted version of the curve in figure~\ref{CVG5}: one that is vertically lowered by 0.109 (cf.~eqs.~(\ref{E28}) and (\ref{E30})).  

If we assume that the radius of the central neutron star of this pulsar has the value $10^6$ cm, i.e.\ $d=1$, then the value $3.38\times10^{12}$ Gauss of its magnetic field $B_0$ (inferred from the conventional formula for magnetic dipole radiation~\cite{Manchester2005}) corresponds, according to the modified version of figure~\ref{CVG5}, to the inclination angle $\alpha=70^\circ$.  This would in turn imply, in conjunction with eq.~(\ref{E29}) and figure~\ref{CVG4}, that the scale factor $j_0$ of the electric current density has the value $4.02\times10^7$ statamp cm$^{-2}$ and the latitudinal direction along which the pulsar is observed is given by $\theta_{P2S}=117^\circ$.  

That the main peaks of the gamma-ray light curve of the Vela pulsar are separated by $\sim0.43$ cycles~\cite{HESS2023} is consistent with 
an inclination angle close to $70^\circ$ (see section~5.1 of~\cite{Ardavan2021}).

\subsection{Characteristics of the source of the Geminga pulsar's emission}
\label{subsec:GemingaSource}

The value of $\kappa$ given in eq.~(\ref{E13}), which applies to a wider range of photon energies than that given in eq.~(\ref{E14}), together with the period ($0.237$~s) and the distance ($190$~pc) of the Geminga pulsar yield
\begin{equation}
{\hat B}_0d^2=2.24\times10^{-1} {\hat\kappa}_{\rm th}^{-1/2}
\label{E31}
\end{equation}
for $\Delta\chi=10^{-3}$ (see eq.~(\ref{E27}) and recall that ${\hat\kappa}_{\rm obs}={\hat\kappa}_{\rm th}$).  The dependence of ${\hat B}_0 d^2$ on $\alpha$ is therefore given by a set of points whose only difference with those plotted in figure~\ref{CVG5} is that their vertical coordinates are shifted upward by $0.065$ (cf.\ eqs.~(\ref{E28}) and  (\ref{E31})).  

Given that the two peaks of the gamma-ray light curve of this pulsar are separated by $\sim0.5$ cycle~\cite{MAGIC2020}, the inclination angle of its central neutron star must be about $60^\circ$ (see section~5.1 of~\cite{Ardavan2021}).  This, together with the modified version of figure~\ref{CVG5}, implies the value ${\hat B}_0\simeq1.22$, i.e.\ $B_0=1.22\times10^{12}$ Gauss, for the magnetic field of the star at its pole (when $d=1$), a value that is not too different from that predicted by the conventional expression for magnetic dipole radiation, i.e.\ $1.63\times10^{12}$ Gauss~\cite{Manchester2005}.  Equation~(\ref{E29}) and figure~\ref{CVG4} respectively yield $j_0=1.12\times10^6$ statamp cm$^{-2}$ and $\theta_{P2S}=90^\circ$ for the scale factor of the electric current density and the colatitude of the observation point in this case.

\section{Concluding remarks}
\label{sec:conclusion}

Even in the most favourable circumstances, spectra of the emissions from relativistic charged particles that are accelerated by synchro-curvature, inverse Compton or magnetic reconnection processes fit the observed data only over disjoint limited ranges of photon energies.  In particular, the constraints set by the $20$ TeV component of the radiation from the Vela pulsar ``provide unprecedented challenges to the state-of-the-art models of HE [high energy] and VHE [very high energy] emission from pulsars''~\cite{HESS2023}.

In contrast, the SED of the tightly focused caustics that are generated by the superluminally moving current sheet in the magnetosphere of a non-aligned neutron star fits the observed spectra of the Crab, Vela and Geminga pulsars over the entire range of photon energies so far detected by Fermi-LAT, MAGIC and HESS from them: over $10^2$ MeV to $20$ TeV.  Not only are these data fitted by a single spectral distribution function but they are also explained by a single emission mechanism, an emission mechanism that has already accounted for a number of other salient features of the radiation received from pulsars: its brightness temperature, polarization, radio spectrum, profile with microstructure and with a phase lag between the radio and gamma-ray peaks~\cite{Ardavan2021,ArdavanHeuristic, ArdavanRadio} and the discrepancy between the energetic requirements of its radio and gamma-ray components~\cite{ArdavanEnergetic}.

Two final remarks concerning the analysis that has led to the present SED are in order:  
\begin{description}
\item[(i)] It is often presumed that the plasma equations used in the numerical simulations of the magnetospheric structure of an oblique rotator should, at the same time, predict any radiation that the resulting structure would be capable of emitting (e.g.~\cite{SpitkovskyA:Oblique,Contopoulos:2012}).  This presumption stems from disregarding the role of boundary conditions in the solution of Maxwell’s equations.  As we have already pointed out, the far-field boundary conditions with which the structure of the pulsar magnetosphere is computed are radically different from the corresponding boundary conditions with which the retarded solution of these equations (i.e.\ the solution describing the radiation from the charges and currents in the magnetosphere) is derived (see section~3 and the last paragraph in section~6 of~\cite{Ardavan2021}).
 
 \item[(ii)] Thickness of the current sheet, which sets a lower limit on the wavelength of the present radiation, is dictated by microphysical processes that are not well understood: the standard Harris solution of the Vlasov-Maxwell equations~\cite{Harris} that is commonly used in analysing a current sheet is not applicable in the present case because the current sheet in the magnetosphere of a non-aligned neutron star moves faster than light and so has no rest frame.  Even in stationary or subluminally moving cases, there is no consensus on whether equilibrium current sheets in realistic geometries have finite or zero thickness~\cite{Klimchuk}.  The fact that the SED described by eq.~(\ref{E5}) yields such good fits to the observed spectra of the pulsars analysed here corroborates the notion that, though necessarily volume-distributed~\cite{BolotovskiiBM:Radbcm}, the magnetospheric current sheet is sufficiently thin to generate gamma rays (see section~4.7 of~\cite{Ardavan2021}).
 
\end{description} 
 
 \section*{acknowledgement} I thank S.\ Campana for his helpful comments on an earlier version of this paper.

\bibliographystyle{plain}

\begin{thebibliography}{}

\end{thebibliography}


\begin{thebibliography}{99}

\bibitem
{Melrose2021} D. B. Melrose, M. Z. Rafat and A. Masterano, {\it Pulsar radio emission mechanisms: a critique, Mon. Not. R. Astron. Soc.} {\bf 500} (2021) 4530--4548.
\bibitem
{HESS2023} The H.E.S.S. Collaboration et al., {\it Discovery of a radiation component from the Vela pulsar reaching 20 teraelectronvolts, Nature Astron.} {\bf 7} (2023), 1341-1350. 
\bibitem
{Ardavan2021} H. Ardavan, {\it Radiation by the superluminally moving current sheet in the magnetosphere of a neutron star, Mon. Not. R. Astron. Soc.} {\bf 507} (2021), 4530--4563.
\bibitem
{SpitkovskyA:Oblique} A. Spitkovsky, {\it Time-dependent force-free pulsar magnetospheres: axisymmetric and oblique rotators, Ap. J.} {\bf 648} (2006), L51--L54.
\bibitem
{Contopoulos:2012} C. Kalapotharakos, I. Contopoulos and D. Kazanas,  {\it The extended pulsar magnetosphere, Mon. Not. R. Astron. Soc.} {\bf 420} (2012), 2793--2798.
\bibitem
{Tchekhovskoy:etal} A. Tchekhovskoy, A. Philippov and A. Spitkovsky, {\it Three-dimensional analytical description of magnetized winds from oblique pulsars, Mon. Not. R.  Astron. Soc.} {\bf 457} (2016), 3384--3395.
\bibitem
{Philippov2022} A. Philippov and M. Kramer, {\it Pulsar magnetospheres and their radiation, Ann. Rev. Astron. Astrophys.} {\bf 60} (2022), 495--558.
\bibitem
{GinzburgVL:vaveaa} V. L. Ginzburg, {\it Vavilov-\v{C}erenkov effect and anomalous Doppler effect in a medium in which the wave phase velocity exceeds the velocity of light in vacuum, Sov. Phys-JETP} {\bf 35} (1972), 92--93.
\bibitem
{BolotovskiiBM:VaveaD} B. M. Bolotovskii and V. L. Ginzburg, {\it The Vavilov-{\v C}erenkov effect and the Doppler effect in the motion of sources with superluminal velocity in vacuum, Sov. Phys. Uspekhi} {\bf 15} (1972), 184--192.
\bibitem
{BolotovskiiBM:Radbcm} B. M. Bolotovskii and V. P. Bykov, {\it Radiation by charges moving faster than light, Sov. Phys. Uspekhi}, {\bf 33} (1990), 477--487.
\bibitem
{ArdavanHeuristic} H. Ardavan, {\it A heuristic account of the radiation by the superluminally moving current sheet in the magnetosphere of a neutron star}, (2022) [astro-ph.HE/2206.02729]. 
\bibitem
{Ardavan2023Crab} H. Ardavan, {\it Congruity of the Crab Pulsar's gamma-ray spectrum with the spectral distribution of tightly focused caustics, Astron. Astrophys.} {\bf 672} (2023), A145.
\bibitem
{Zanin2017} R. Zanin, {\it The Crab pulsar at VHE, Eur. Phys. J. Web Conf.} {\bf 136} (2017), 03003.
\bibitem
{Stamnes1986} J. J. Stamnes, {\it Waves in Focal Regions}, A. Hilger, Bristol, U.K. (1986).
\bibitem
{Olver} F. W. J. Olver, D. W. Lozier, R. F. Boisvert and C. W. Clark, {\it NIST Handbook of Mathematical Functions}, Cambridge Univ. Press, Cambridge, U. K. (2010).
\bibitem
{Abdo2010} A. A. Abdo, M. Ackermann, M. Ajello et al. {\it Fermi Large Area Telescope observations of the Crab pulsar and nebula, Astrophys. J.} {\bf 708} (2010), 1254--1267.
\bibitem
{Aleksic2011} J. Aleksi\'{c}, E. A. Alvarez, L. A. Antonelli et al. {\it Observations of the Crab pulsar between 25 and 100 GeV with the MAGIC I telescope, Astrophys. J.} {\bf 742} (2011), article id. 43.
\bibitem
{Ansoldi2016} S. Ansoldi, L. A. Antonelli, P. Antoranz et al. {\it Teraelectronvolt pulsed emission from the Crab pulsar detected by MAGIC, Astron. Astrophys.} {\bf 585} (2016), A133.
\bibitem
{Abdo2013} A. A. Abdo, M. Ajello, A. Allafort et al. {\it The {S}econd {F}ermi {L}arge {A}rea {T}elescope {C}atalog of {G}amma-ray {P}ulsars, Astrophys. J. Suppl.} {\bf 208} (2013), 17--76.
\bibitem
{MAGIC2020} The MAGIC Collaboration et al. {\it Detection of the Geminga pulsar with MAGIC hints at power-law tail emission beyond 15 GeV, Astron. Astrophys.} {\bf 643} (2020), L14.
\bibitem
{Manchester2005} R. N. Manchester, G. B. Hobbs, A. Teoh and M. Hobbs, {\it The Australia Telescope National Facility Pulsar Catalogue, Astrophys. J.} {\bf 129} (2005), 1993--2006.
\bibitem
{ArdavanRadio} H. Ardavan, {Radio spectra of pulsars fitted with the spectral distribution function of the emission from their current sheet}, (2024) [astro-ph.HE/2402.09913].
\bibitem
{ArdavanEnergetic} H. Ardavan, {\it Energetic requirements of the gamma-ray emission from pulsars: A nonparametric analysis of the data in the Fermi-LAT 12-Year Catalog, J.~High~Energy~Astrophys.} {\bf 37} (2023), 62--70.
\bibitem
{Harris} E. G. Harris, {\it On a plasma sheath separating regions of oppositely directed magnetic field, Nuovo Cim.} {\bf 23} (1962), 115--121.
\bibitem
{Klimchuk} J. A. Klimchuk, J. E. Leake, L. K. S. Daldorff and C. D. Johnston, {\it The thickness of current sheets and implications for coronal heating, Front. Phys.} {\bf 11} (2023), 1198194





\end{thebibliography}

\end{document}